\documentclass[twocolumn,showpacs,preprintnumbers,amsmath,amssymb]{revtex4}
\catcode`\@=11
\def\eqlt{\mathrel{\mathpalette\@vereq<}}  
\def\eqgt{\mathrel{\mathpalette\@vereq>}}  
\def\@vereq#1#2{\lower2.5pt\vbox{\baselineskip0pt \lineskip-.5pt
 \ialign{$\m@th#1\hfil##\hfil$\crcr#2\crcr{=}\crcr}}}

\newcommand{\simle}{\ \raise.3ex\hbox{$<$}\kern-0.8em\lower.7ex\hbox{$\sim$}\ }
\newcommand{\simge}{\ \raise.3ex\hbox{$>$}\kern-0.8em\lower.7ex\hbox{$\sim$}\ }

\begin{document}
\title {New Type of Degenerate Quantum Spin Phase}  
\author { Masatoshi Imada$^{1,2}$, Takahiro Mizusaki$^{3}$  and Shinji Watanabe$^{1}$}
\address {$^1$Institute for Solid State Physics, University of Tokyo,  
Kashiwanoha, Kashiwa, 277-8581, Japan}
\address {$^2$PRESTO, Japan Science and Technology Corporation}  
\address {$^3$Institute of Natural Sciences, Senshu University, Higashimita, 
Tama, Kawasaki, 214-8580, Japan}  

\begin{abstract} 

Correlated electrons often crystalize to the Mott insulator usually with some magnetic orders, whereas the ``quantum spin liquid" has been a long-sought issue.  
We report numerical evidences that a nonmagnetic insulating (NMI) phase gets stabilized near the Mott transition with remarkable properties:
 The 2D Mott insulators on geometrically frustrated lattices contain a phase with gapless spin excitations and degeneracy of the ground state in the whole Brillouin zone of the total momentum.  It has an interpretation for an
unexplored type of a quantum liquid.
The present concept is useful in analyzing a variety of experimental results in frustrated magnets including organic BEDT-TTF compounds.
\end{abstract}
\maketitle
Among various insulating states, those caused by electronic Coulomb correlation effects, called 
the Mott insulator, show many remarkable phenomena such as high-Tc superconductivity and colossal magnetoresistance near it~\cite{Mott,RMP}. However, it has also been an issue of long debate whether the Mott insulator has its own identity distinguished and adiabatically separated from other insulators like the band insulator.
This is because the Mott insulator in most cases shows symmetry breakings such as antiferromagnetic order or dimerization, where the resultant folding of the Brillouin zone makes the band full and such insulators difficult to distinguish from the band insulators
because of the adiabatic continuity.

Except in one dimension, the possibility of the inherent Mott insulator without conventional orders has been a long-sought challenge.  
The Mott insulator on the triangular lattice represented by the Heisenberg spin systems was proposed as a candidate, where the spins quantum mechanically melt against spin solidifications~\cite{Anderson}.  Although, the triangular Heisenberg system itself has been controversial and argued theoretically to show an antiferromagnetic (AF) order~\cite{Lhuillier}, intensive studies on geometrical frustration effects have been stimulated.  

 In this letter, we show a numerical evidence for the existence of a new type of inherent Mott insulator near the Mott transition; singlet ground state with unusual degeneracy in the total momentum accompanied by gapless and dispersionless spin excitations.   

Recently extensive experimental studies on frustrated quantum magnets such as those on triangular, Kagome, spinel and pyrochlore lattices have been performed~\cite{Ramirez,Greedan}. 
They tend to show 
suppressions of magnetic orderings with large residual entropy with
a gapless liquid feature for quasi 2D systems or ``spin glass-like" behavior in 3D even for disorder-free compounds. These gapless and degenerate behaviors wait for a consistent theoretical understanding.
Our present results offer a useful underlying concept for the understanding of the puzzling feature.    

\begin{figure}
$$ \psboxscaled{300}{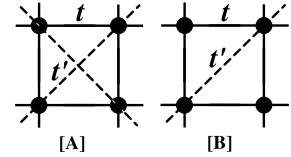} $$
\caption{Lattice structure of geometrically frustrated lattices [A] on a square lattice and [B] on an anisotropic triangular lattice.  The nearest- and next-nearest-neighbor transfers are denoted by $t$ and $t'$, respectively}
\label{Fig1}
\end{figure}

To get an insight into this issue, our study focuses on the Hubbard model on two-dimensional frustrated lattices.  The Hamiltonian by the standard notation reads 

\begin{eqnarray}  H&=&-\sum_{\langle i,j \rangle ,\sigma}t   
\left(c_{i\sigma}^{\dagger}c_{j\sigma}+H.c.\right)
  + \sum_{\langle k,l \rangle,\sigma}t' \left(c_{k\sigma}^{\dagger}c_{l\sigma}+{\rm H.c.}\right)
\nonumber\\ 
 & &+U\sum_{i=1}^{N} \left(n_{i\uparrow}-\frac{1}{2}\right)  
\left(n_{i\downarrow}-\frac{1}{2}\right) 
\label{Hamiltonian}  
\end{eqnarray} 
on a $N$-site square lattice with a nearest neighbor ($t$) and diagonal next-nearest neighbor ($t'$) transfer integrals in the configurations [A] and [B] illustrated in Fig.~\ref{Fig1}. The energy unit is taken by $t$.  The insulating phase is stabilized when the on-site Coulomb repulsion $U$ is large enough, 
while the metallic phase appears at small $U$ when $t'$ is nonzero as we will show.  

\input psbox.tex

\begin{figure}
$$ \psboxscaled{500}{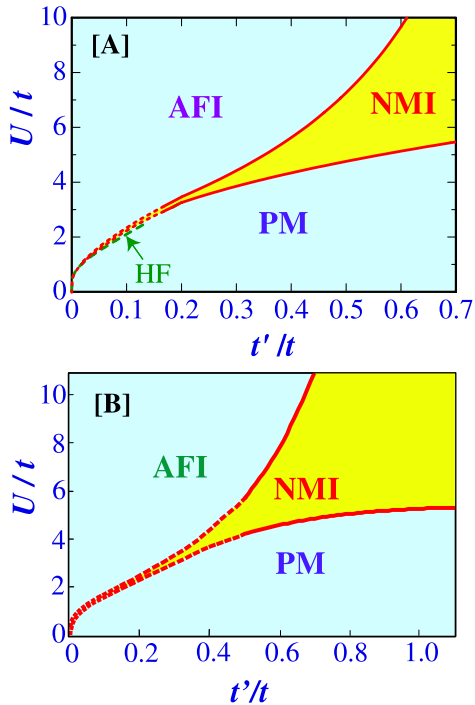} $$
\caption{(color) Phase diagrams of the Hubbard models with the lattice structure illustraterd in Fig. 1 [A] and [B] in the parameter space of $U$ scaled by $t$, and the frustration parameter $t'/t$.  AFI, PM, and NMI represent the antiferromagnetic insulating, paramagnetic metallic and nonmagnetic insulating phases, respectively. The dashed curve indicated by "HF" in the upper panel is the Hartree Fock result~\cite{Vollhardt}}
\label{Fig2}
\end{figure}

The model requires accurate and unbiased theoretical calculations because of large fluctuation effects expected from the low dimensionality of space and the geometrical frustration effects due to nonzero $t'$.  
Recently, the path-integral renormalization group (PIRG) method~\cite{Kashima-Imada} opened a way of numerically studying the models with the frustration effects more thoroughly without the negative sign problem and without relying on the Monte Carlo sampling. The efficiency of the method was established through a number of applications~\cite{Kashima,Morita,Noda,MizusakiN}. By using the PIRG method, the existence of a nonmagnetic insulator (NMI) near the metal-insulator transition boundaries was reportred~\cite{Kashima,Morita} on the two-dimensional frustrated Hubbard model on lattices [A] and [B].  
In Fig.~\ref{Fig2}, the phase diagrams are illustrated~\cite{Kashima,Morita}.  
The phase diagrams show quantum melting of spin orders at higher $U$ than the Mott transition. This new aspect is ascribed to enhanced charge fluctuations and increasing double occupation near the Mott transition, which cannot be studied in the Heisenberg models.  However, the nature of the resultant NMI in 2D remained totally unclear. We reveal a couple of remarkable features of this phase in this letter.  

We study the excitation spectra calculated by the PIRG method. 
More specifically, by following the PIRG method~\cite{Kashima-Imada}, the wavefunction with the form
$|\Psi\rangle=\sum_{j=1}^{L}c_{j}|\phi_j\rangle$
is numerically optimized so that the energy estimate $\langle \Psi \vert H \vert \Psi \rangle /\langle \Psi \vert \Psi \rangle$ becomes the lowest among choices of arbitrary nonorthogonal Slater basis functions $\vert \phi_j \rangle $ and the coefficients $c_j$.  The optimization of $\vert \phi_j \rangle $ and $c_j$ are achieved by repeated operations of $\exp[-\tau H]$ to $\Psi$. 
The convergence to the true lowest energy state is obtained after the systematic increase of $L$. 
In the present work, we have improved the original PIRG algorithm to obtain the lowest energy state with specific quantum numbers. Such quantum number projection can be performed through a rotation of the state $\vert \Psi \rangle$ with an angle $\phi$ in the spin space using the rotation operator ${\cal R}(\phi)$ and by a spatial translation with a shift of ${\bf r}$ by the translation operator ${\cal L}({\bf r})$ in the process of PIRG.  A state with specific quantum numbers $S$ and ${\bf k}$ is obtained from the weighted integration as 
$\vert \Phi (S,{\bf k}) \rangle = \int d\phi W_R(S,\phi)R(\phi) \sum_{\bf r} W_L({\bf k},{\bf r}){\cal L}({\bf r})\vert \Psi \rangle$.
Here the weights $W_R$ and $W_L$ are chosen to specify the quantum numbers $S$ and ${\bf k}$.  For example, $W_L({\bf k},{\bf r})=\exp[i{\bf k} \cdot {\bf r}]$.
Since the $z$ component of the total spin, $S^z$, is fixed in $\vert \Psi \rangle$, the integral over the Euler angle is reduced to that over a single variable $\phi$. 
This quantum number projection procedure also improves the accuracy of the energy estimate substantially. The accuracy of this method was carefully examined and confirmed in many examples. For example, all the ground states and excitation spectra studied on $4\times4$ lattice show roughly 4 digit accuracy in comparison with the available exact results.

\begin{figure}
$$ \psboxscaled{300}{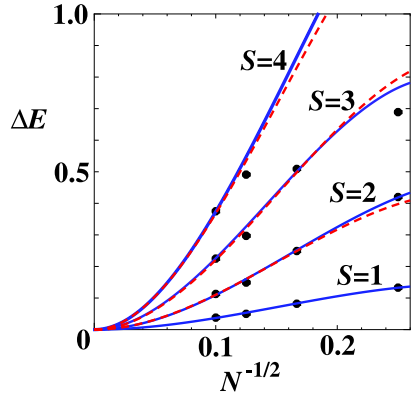} $$
\caption{(color) Size scalings of the energy gaps for total spin $S$=1,2,3 and 4 in the AFI phase ($U=4,t=1,t'=0$).  The solid curves are fitting by the form (2) and the dashed curves illustrate curves obtained from the $S=0$ fitting multiplied with the factor $4S(S+1)/3$.}
\label{Fig3}
\end{figure}

Before reporting proprties in the NMI phase, we first discuss our PIRG result in the AFI phase.   
Here, the finite size gap for a $\ell \times \ell$ system in the chiral perturbation theory~\cite{Hasenfratz-Niedermayer} in the form
\begin{eqnarray}  
\Delta E= \frac{c^2}{\rho \ell^2}[1-\frac{3.900265c}{4\pi\rho \ell}+O(\frac{1}{\ell^2})]
\label{Spinwave}  
\end{eqnarray} 
 are fitted with the calculated results in Fig.~\ref{Fig3}. The finite-size gap nicely follows the form (\ref{Spinwave}) in the AFI phase. For example, at $U=4, t=1$ and $t'=0$, the fitting in Fig.~\ref{Fig3} shows the spin wave velocity $c\sim 0.67$ and the spin stiffness $\rho \sim 1/7.1c$, which are equivalent to the estimate of the Heisenberg model at the exchange coupling $J=0.42$ in the spin wave theory. 
The fitted values of $c$ and $J$ well reproduce the previous estimates (ex. $J \sim 0.4$) obtained from the susceptibility and the staggered magnetizations~\cite{Hirsch}.
The excitations in the AFI phase well satisfy the tower structure of the low-energy excitation spectra based on the nonlinear sigma model description.  

\input psbox.tex

\begin{figure}
$$ \psboxscaled{300}{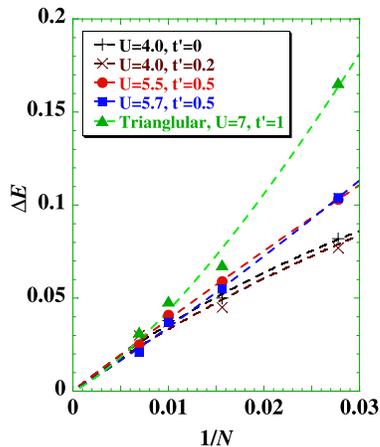} $$
\caption{(color) Size scalings of the $S=1$ excitation gaps for several choices of parameters 
in the NMI phase.  The triangles show the case of the model [B] while others are for the model [A].  The dashed curves are fittings to Eq.(2). The circles, squares and triangles are results in the NMI phase. }
\label{Fig4}
\end{figure}

Now in the NMI phase, typical system size dependences of the spin excitation gap $\Delta E$ between the singlet ground state and the lowest triplet state are shown in Fig.~\ref{Fig4}.  All of the data points in Fig.~\ref{Fig4} 
indicate that the triplet excitations become gapless in the thermodynamic limit. 
The gap appears to be scaled asymptotically with the inverse system size $N^{-1}$, namely $\Delta E\sim \zeta/N$.  The gapless feature shares actually some similarity to the behavior in the AFI phase. 
 However detailed comparison clarifies a crucial difference as we will show later.  The fitting of the data in the NMI phase to the form (2) gives unphysical values such as $c>1.5$.
We note that the uniform magnetic susceptibility is given by $2/3\zeta$. Therefore, the present data imply that the uniform susceptibility becomes a nonzero constant.

\input psbox.tex

\begin{figure}
$$ \psboxscaled{500}{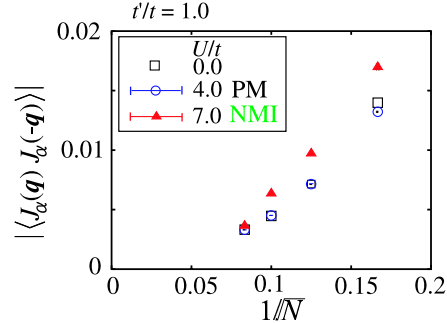} $$
\caption{(color) Size scalings of the staggered flux correlations for the model [B] at $t'/t=1.0$}
\label{Fig5}
\end{figure}

Except in 1D systems, the present result is the first numerical evidence by unbiased calculations for the existence of gapless excitations without apparent long-ranged order in the Mott insulator.
Although a tiny order cannot be excluded if it is beyond our numerical accuracy, in the present NMI phase, the absence of various symmetry breakings including the AF order has already been shown in the model [B]~\cite{Morita}. In the model [A], the size scaling suggests the absence of AF order at any wavevector as well~\cite{Kashima-Imada}. As well as dimer and  plaquette singlet orders, s- and d-density waves are also numerically shown to be unlikely for four types of correlations probed by
$
C_{\alpha}({\bf q})=
\left|
\langle J_{\alpha}({\bf q})J_{\alpha}^{\dagger}({\bf q}) \rangle
\right|
$, 
$
J_{\alpha}({\bf q})=\frac{1}{N}\sum_{{\bf k},\sigma}
c^{\dagger}_{{\bf k},\sigma}c_{{\bf k}+{\bf q},\sigma}f_{\alpha}({\bf k})
$,
with 
$f_{1}({\bf k})=\cos(k_{x})+\cos(k_{y})$, 
$f_{2}({\bf k})=\cos(k_{x})-\cos(k_{y})$, 
$f_{3}({\bf k})=2\cos(k_{x})\cos(k_{y})$, 
and 
$f_{4}({\bf k})=2\sin(k_{x})\sin(k_{y})$. 
%
Fig.~\ref{Fig5} in the example of $\alpha=2$ clarifies the correlation of the d-density wave (namely, the staggered flux)~\cite{Affleck}.

The dispersions of the $S=1$ excitations show dramatic difference between the AFI and NMI phases.  
Here the dispersions are given from the lowest energy states with specified momenta, $E({\bf k})$ calculated from the spin-momentum resolved PIRG.  In the AFI phase, the dispersion is essentially described by the spin-wave spectrum of the form similar to $\Delta E(k) =4J\sqrt{1-\gamma_k^2}$ with $\gamma_k=\frac{1}{2}(\cos(k_x)+\cos(k_y))$ for the spin wave theory of the Heisenberg model, but modified because of finite $U$.  
The calculated dispersion width ($\sim 1.5$) is comparable to the estimate of the spin-wave dispersion at $J=0.4$.

\begin{figure}
$$ \psboxscaled{300}{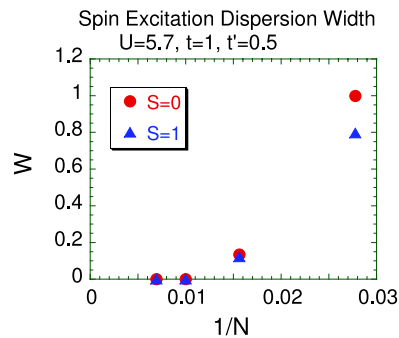} $$
\caption{(color) Size scalings of the dispersion widths for $S=0$ and $S=1$ excitations in the NMI phase for the model [A].}
\label{Fig6}
\end{figure}

In marked contrast, the $S=1$ dispersion in the NMI phase has strong and monotonic system size dependence as in Fig.~\ref{Fig6}.  For systems larger than $8 \times 8$ lattice, the dispersion surprisingly becomes vanishingly small.   The size dependence shows very quick collapse of the dispersion with increasing system size and may not be fitted by a power of the inverse system size as in the single-particle Stoner excitations in metals.   Such a flat dispersion is observed solely in the NMI phase irrespective of the models (namely commonly seen in the models [A] and [B]).  

\input psbox.tex

The presence of such degenerate excitations well accounts for the quantum melting of simple translational symmetry breakings including the AF long-ranged order, because any type of long-ranged order in the two-dimensional systems is destroyed when the excitation becomes flatter than $\Delta E(k)=k^2$. This is because of the infrared divergence in the form $\int k^{d-1}dk\frac{1}{\Delta E(k)}$ with $d=2$ when one calculates the fluctuations around the order.  

In addition to $S=1$ excitations, the total singlet state ($S=0$) at any total momentum ${\bf k}$ also shows degenerate structure in the ground state for larger system size as in Fig.~\ref{Fig6}. The dispersion vanishes in the NMI phase in the model [A] as well as in [B].



The present excitation spectra show the following double-hierarchy structure: The ground states are degenerate within the total spin $S=0$ sector among different total momenta; the dispersion quickly collapses with increasing system size.  Another degeneracy, the gapless spin excitation among different total spins, emerges much slowly with increasing system size.

Here we discuss a possible interpretation of the properties. 
The vanishing gap may allow an interpretation that the dynamical singlet bonds filling the whole lattice contained in the ground state wavefunction has a nonzero weight of distribution at vanishingly small singlet binding energy, thus the distribution of the singlets over long distance with weights of presumable power law decay as in a variational long-ranged RVB wavefunction~\cite{Hsu}. 

  The collapse of dispersion shows that a ground state is degenerate with other ground states obtained by spatial translations and they have vanishing off-diagonal Hamiltonian-matrix elements each other.  This degeneracy implies the orthogonality catastrophe under spatial translations.
Although the translational symmetry is retained in the original Hamiltonian, the present orthogonality supports an unconventional symmetry breaking in the whole Brillouin zone of the total momenta.
The excitations cannot be coherent propagative modes but are localized because of the lack of dispersion.  Coherent (or Stoner-type) spinon excitations with a spinon Fermi surface do not seem to explain the very quick collapse of the dispersion with the increasing system size as well.
From the long-ranged RVB picture, one can argue that an unbound singlet does not coherently propagate because of scattering by other dynamical singlets. 

This nonmagnetic insulator appears to be stabilized simply due to the Umklapp scattering~\cite{Rice}. The continuum of degenerate excitations within the singlet sector, which is similar to the present results, but in the presence of the spin gap was proposed in the Kagome and pyrochlore lattices based on small cluster studies~\cite{Chalker}. The possible symmetry breaking from degenerate singlet states was also examined on a pyrochlore lattice~\cite{Harris,Tsunetsugu}, while the spin excitations were again argued to be gapful in contrast to the present results.  In our results, the degeneracy becomes clear only at larger system sizes than those in these studies.  

We briefly discuss experimental implications of the present new quantum phase.
Recent results by Shimizu {\it et al.}~\cite{Kanoda} on $\kappa$-(ET)$_2$Cu$_2$(CN)$_3$  appear to show an experimental realization of the quantum phase we have discussed in the present work.  In fact this compound can be modelled by a single band Hubbard model on nearly right triangular lattice near the Mott transition.  The NMR relaxation rate  
shows the nonmagnetic and gapless nature retained even at low temperatures ($\sim 0.2$K) and suggests the present quantum phase category. Another organic compound also shows a similar behavior~\cite{Kato}     

Systems with Kagome-like structure, $^3$He on graphite~\cite{Ishida} and volborthite Cu$_3$V$_2$O$_7$(OH)$_2$$\cdot$2H$_2$O~\cite{Hiroi} show nonmagnetic and gapless behaviors.  
On the other hand, glass-like transitions are seen in 3D systems, typically in pyrochlore compounds as $R_2$Mo$_2$O$_7$ with $R=$Er, Ho, Y, Dy, and Tb~\cite{Taguchi} and in fcc structure, Sr$_2$CaReO$_6$~\cite{Wiebe}.  It is remarkable that the glass behavior appears to occur without introducing the disorder.  
Although the lattice structure, dimensionality and local moments have a diversity, many frustrated magnets show gapless and incoherent(glassy) behavior. The present result on gapless and strongly degenerate structure emerging without quenched randomness offers a consistent concept with this universal trends. It would be an extremely interesting and open theoretical issue whether the present unusual translational symmetry breaking leads to such a glass phase at $T=0$ in 2D.

In summary, we have theoretically clarified the existence of a new degenerate quantum spin phase in the Mott insulator under geometrical frustration effects.  The phase has gapless spin excitations from the degenerate ground states, and furthermore the dispersionless modes are found in all the spin sector.  This may cause an unusual translational symmetry breaking. 
Recent experimental findings, though diverse, tend to show a relevance of this phase in disorder-free and frustrated systems.      

A part of the computation was done at the supercomputer center in ISSP, University of Tokyo.   

\end{document}